# THE MECHANICAL DESIGN FOR THE DARHT-II DOWNSTREAM BEAM TRANSPORT LINE*

G. A. Westenskow, L. R. Bertolini, P.T. Duffy, A. C. Paul
Lawrence Livermore National Laboratory, Livermore, CA 94550 USA

*Abstract.*

This paper describes the mechanical design of the downstream beam transport line for the second axis of the Dual Axis Radiographic Hydrodynamic Test (DARHT II) Facility. The DARHT-II project is a collaboration between LANL, LBNL and LLNL. DARHT II is a 20-MeV, 2000-Amperes, 2-μsec linear induction accelerator designed to generate short bursts of x-rays for the purpose of radiographing dense objects. The downstream beam transport line is approximately 20-meter long region extending from the end of the accelerator to the bremsstrahlung target. Within this proposed transport line there are 15 conventional solenoid, quadrupole and dipole magnets; as well as several speciality magnets, which transport and focus the beam to the target and to the beam dumps. There are two high power beam dumps, which are designed to absorb 80-kJ per pulse during accelerator start-up and operation. Aspects of the mechanical design of these elements are presented.

## 1 INTRODUCTION

We are working on the engineering design of the downstream beam transport components of the DARHT II Accelerator [1]. Beam transport studies for this design have been performed [2]. Figure 1 shows the proposed layout for the elements in the system during early commissioning of the downstream components. The beamline from the exit of the accelerator to the target is about 20 meters long. In the accelerator the pulse length is about 2 μsec. However, only four short (20 to 100 nsec) pulses separated by about 600 nsec are desired at the bremsstrahlung target. The function of the kicker system is to "kick" four shorter pulses out from the main 2-μsec. The kicker includes a bias dipole operated so that the non-kicked parts are deflected off the main line into the main beam dump, while the kicked pulses are sent straight ahead. Focusing elements between the kicker and the septum would complicate operation. Therefore, to achieve a narrow beam waist at the septum, solenoid S3 must "throw" a waist to the septum. The first 3 meters of beamline allow the beam to expand from its 5-mm matched radius in the accelerator to 2.25-cm at solenoid S3. The system is designed to have a 20% energy acceptance to transport the main beam and most of the leading and falling edge of the pulse exiting the accelerator. The proposed system using a quadrupole magnet [2] allows for a larger beam pipe radius than the more conventional septum dipole magnet studied earlier. This increases the energy acceptance of the transport line to the main beam dump.

Work on the kicker system has been described elsewhere[3]. After the septum, there are four Collins style quadrupole magnets to restore the beam to a round profile. Experience has shown ejecta material from the target can travel the length of the accelerator. To keep this material from reaching the injector and accelerator cells we have included a rotating wheel debris blocker that does not allow direct line-of-sight between the target and the upstream elements for times shortly after the pulse.

Finally the beam will be pinched to a tight focus at the target to provide an intense spot of x-rays for radiographic purposes. Work on the target is also presented in these proceedings [4].

## 2 TRANSPORT ELEMENTS

The magnets within the DARHT II transport line are all water-cooled conventional dc electromagnets (except the bias dipole and the kicker sextupole corrector). The magnets are listed in Table 1.

The transport solenoids have external iron shrouds with water-cooled copper coils. Solenoid coils are wound into individual two-layer "pancake" coils. Each magnet has an even number of these pancake coils. The pancakes are installed in an A-B-A-B orientation to minimize axial field errors. The inside diameters of the solenoid coils are sized large enough to fit over the outside diameter of the beam tube flanges.

The septum quadrupole and dipole magnets have solid iron cores with water-cooled copper coils. The Septum Quadrupole Magnet is a four-piece, solid-core construction. The Collins Quadrupole Magnets are two-piece solid cores with non-magnetic support. The dipole magnet is a three-piece, solid-core, "C" magnet.

The alignment requirements for the transport magnets are $\pm 0.4$-mm positional tolerance and $\pm 3$-mrad angular tolerance. Steering corrector coils will be installed under many of the magnetic elements to allow for small alignment errors or stray magnetic fields.

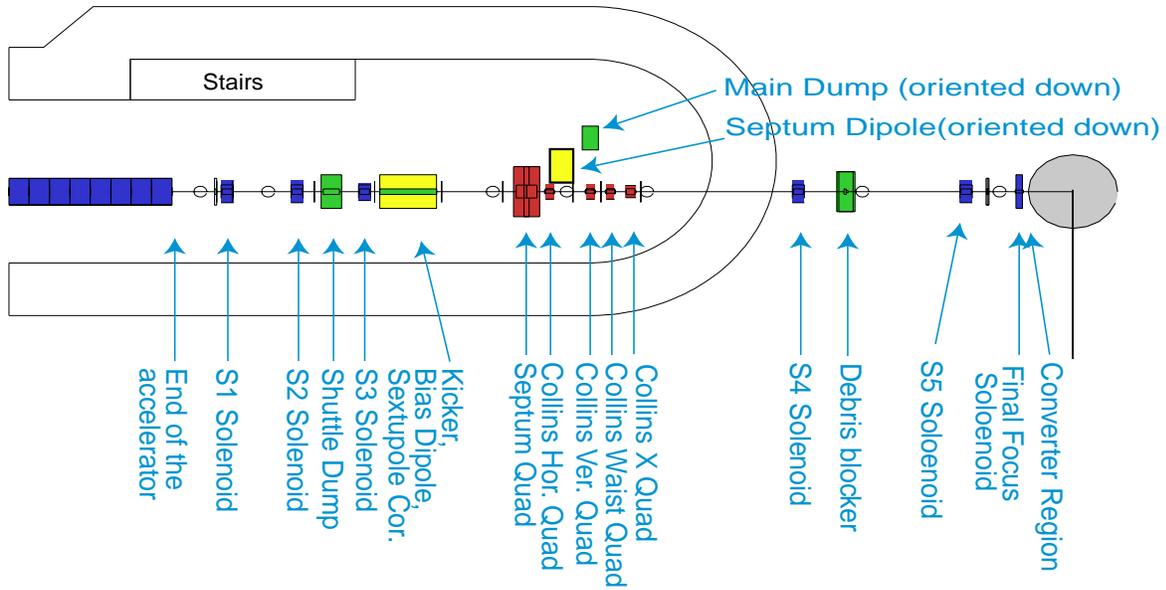

Figure 1: Layout of the transport elements.

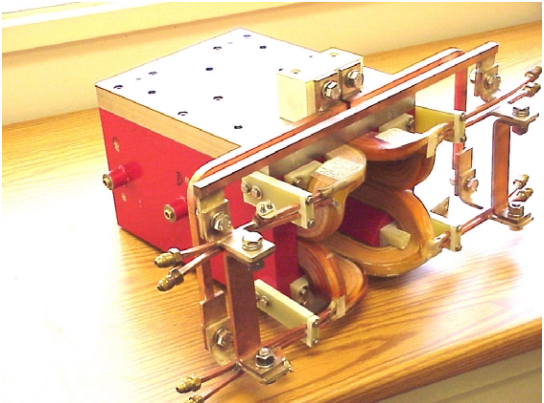

Figure 2: Picture of a Collins Quadrupole.

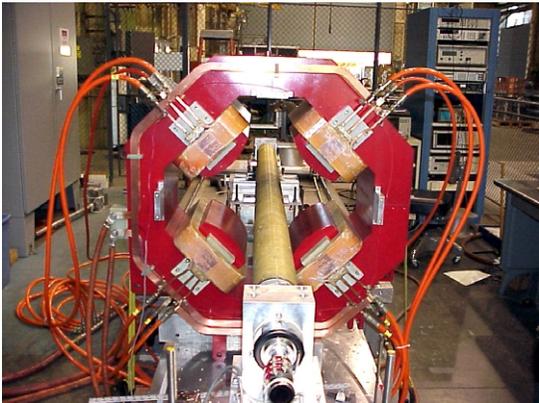

Figure 3: Picture of the Septum Quadrupole being characterised during a rotating coil test.

**Transport Elements in the Downstream Beamline.**

| Magnet Type | Magnet Name | Max. Field (kG) or Gradient (kG/m) | Bore or Gap (cm) |
|---|---|---|---|
| Solenoid | S1 | 8 | 27 |
| Solenoid | S2 | 8 | 27 |
| Solenoid | S3 | 2.5 | 27 |
| Dipole | Bias Dipole | 0.009 | 41 |
| Pulsed Dipole | Kicker | -0.009 (equivalent) | 12.8 |
| Sextupole | Sextupole Corrector | 16 gauss @20.5 cm radius | 41 |
| Quadrupole | Septum Quadrupole | 8.0 | 38 |
| Quadrupole | Collins-H | 10.0 | 12 |
| Quadrupole | Collins-V | 10.0 | 12 |
| Quadrupole | Collins-W | 10.0 | 12 |
| Quadrupole | Collins-X | 10.0 | 12 |
| Solenoid | S4 | 2.5 | 27 |
| Solenoid | S5 | 2.5 | 27 |
| Solenoid | Final Focus Solenoid | ~5 | ~13 |
| Dipole | Septum Dipole | 1.0 | 16 |

Table 1: Magnet specifications

## 3 VACUUM SYSTEM

The vacuum chambers for the DARHT II Transport Line are circular beam pipes constructed from 304L stainless steel. The region from the end of the accelerator through the septum has a 16-cm bore diameter. From the septum to the target, the bore diameter is reduced to 9.55 cm. The vacuum seals are made with conflat style knife-edge flanges with annealed copper gaskets. The use of all-metal seals is driven by the potential requirement to *in situ* bake the transport vacuum system. *In situ* bake-out may be required to minimize adsorbed gas on the beam tube walls, which may be desorbed by beam halo scraping the walls. The vacuum design requirement for the transport line is $10^{-7}$ Torr pressure.

Figure 4 shows a side view of the septum vacuum chamber. The chamber resides in the region were the beamline splits between the line going to the target and the line going to the main dump. The chamber is formed by two aluminium halves that are then welded together at the midplane.

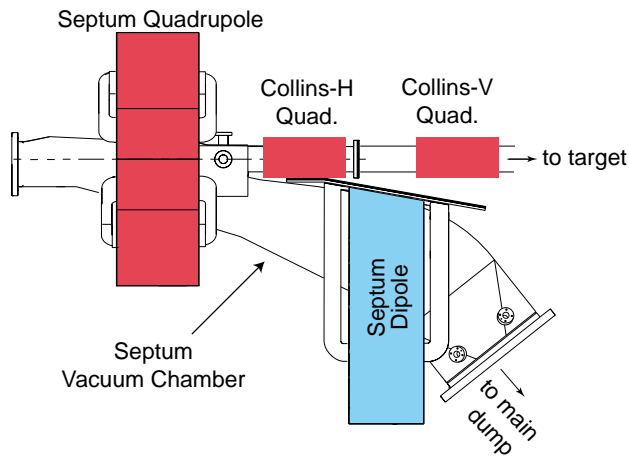

Figure 4: Arrangement of the transport elements around the septum. Horizontal view. The main part of the pulse enters the Septum Quadrupole off-axis and is bent into the Septum Dipole, it is then bent further into the main dump. The kicked portion goes straight ahead though the Collins Quadrupoles.

## 4 BEAM DUMPS

There are two beam dumps included in the DARHT II downstream transport system; a main beam dump, and a shuttle dump. The purpose of the shuttle dump is to allow accelerator operations while personnel are working in the target area outside the accelerator hall. The shuttle dump will have a composite absorber, made up of a 3-inch thick graphite block, backed by 12-inches of tungsten. There will be additional shielding surrounding the beam stop to absorb radiation.

The main beam dump absorbs the portion of the beam that is not deflected by the kicker system. The normal horizontal beam size at the main beam dump is 8 cm. However, the start-up parameters for the beam will not be well known. We must therefore provide some safety margin. First consideration is to keep the instantaneous temperature (temperature at the end of the 2-μsec pulse) of the impact area below the damage point for the material. At 1 pulse per minute repetition rate we can manage the average temperature increase. We also desire to keep the neutron yield low to minimize activation of components and simplify radiation shielding. The construction of the beam dump must be compatible with high vacuum as explained in the previous section.

## 5 DIAGNOSTICS

Throughout the beamline there are beam position monitors (BPMs) to measure the location and angle of trajectory of the beam. The BPMs mount between the flanges of adjacent transport beam tubes. The accurate transverse location of the BPMs is critical to the operation of the transport line and it is their positional requirements, which set the alignment tolerances for the beam line vacuum system. To satisfy the BPM requirements, the transport line beam tubes must be aligned to within 0.2 mm offset. Each beam tube will be manufactured with fiducials, which allow the survey crew to measure and position external to the centerline.

During commissioning rather than going into the "target" region, we will have beam diagnostics on the straight ahead beamline. They will include a spectrometer for measuring beam energy, and a "pepper-pot" for measuring beam emittance.

The septum vacuum chamber has several ports near the septum region to allow characterisation of the plasma generated during pulses when the beam is spilt on nearby walls.